\documentclass[sigconf, language=russian, language=english]{acmart}

\AtBeginDocument{%
  }

\acmConference[Draft]{Manuscript}{May}{2023}

\begin{document}

\title{Detecting Human Rights Violations on Social Media during Russia-Ukraine War}

\author{Poli Nemkova}
\email{poli.nemkova@unt.edu}
\affiliation{
  \institution{University of North Texas}
  \city{Denton}
  \state{Texas}
  \country {USA}
}

\author{Solomon Ubani}
\affiliation{
  \institution{University of North Texas}
  \city{Denton}
  \state{Texas}
  \country {USA}
}
\email{solomonubani@my.unt.edu}

\author{Suleyman Olcay Polat}
\affiliation{
  \institution{University of North Texas}
  \city{Denton}
  \state{Texas}
  \country {USA}
}
\email{suleymanolcaypolat@my.unt.edu}

\author{Nayeon Kim}
\affiliation{
  \institution{University of North Texas}
  \city{Denton}
  \state{Texas}
  \country {USA}
}
\email {nayeonkim3@my.unt.edu}

\author{Rodney D. Nielsen}
\affiliation{
  \institution{University of North Texas}
  \city{Denton}
  \state{Texas}
  \country {USA}
}
\email{rodney.nielsen@unt.edu}

\renewcommand{\shortauthors}{Nemkova et al.}

\begin{abstract}
 The present-day Russia-Ukraine military conflict has exposed the pivotal role of social media in enabling the transparent and unbridled sharing of information directly from the frontlines. In conflict zones where freedom of expression is constrained and information warfare is pervasive, social media has emerged as an indispensable lifeline. Anonymous social media platforms, as publicly available sources for disseminating war-related information, have the potential to serve as effective instruments for monitoring and documenting Human Rights Violations (HRV). Our research focuses on the analysis of data from Telegram\footnote{https://telegram.org/}, the leading social media platform for reading independent news in post-Soviet regions\footnote{https://www.wsj.com/articles/telegram-thrives-amid-russias-media-crackdown-11647595800}. We gathered a dataset of posts sampled from 95 public Telegram channels that cover politics and war news, which we have utilized to identify potential occurrences of HRV. Employing a mBERT-based text classifier, we have conducted an analysis to detect any mentions of HRV in the Telegram data. Our final approach yielded an $F_2$ score of 0.71 for HRV detection, representing an improvement of 0.38 over the multilingual BERT base model. We release two datasets that contains Telegram posts: (1) large corpus with over 2.3 millions posts and (2) annotated at the sentence-level dataset to indicate HRVs. The Telegram posts are in the context of the Russia-Ukraine war. We posit that our findings hold significant implications for NGOs, governments, and researchers by providing a means to detect and document possible human rights violations.
\end{abstract}

\ccsdesc[500]{Computing methodologies~Natural language processing}
\ccsdesc[500]{Computing methodologies~Machine learning}
\ccsdesc[500]{Computing methodologies~Information extraction}

\keywords{Natural Language Processing, Text Classification, Social Media, Human Rights Violations, Transformers, ChatGPT, Russian}

\begin{teaserfigure}
  \includegraphics[width=\textwidth]{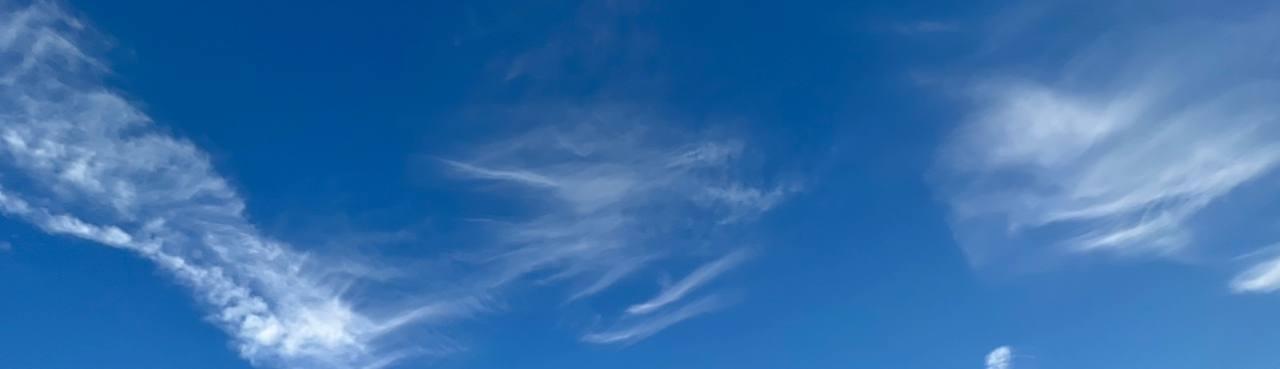}
  \caption{Peaceful Sky, Eastern Europe.}
  \label{fig:teaser}
\end{teaserfigure}

\received{15 March 2023}

\maketitle

\section{Introduction}

The Russia-Ukraine war has been a concern for Human Rights Violations (HRVs) since its beginning, with  reports of atrocities committed towards civilians and soldiers alike.  As early as two months after the war began, the Office of the United Nations High Commissioner for Human Rights (OHCHR) estimated nearly six thousand civilian casualties \cite{bachelet2022atrocity}. Another two months later, a formal report from the same organization verified an increase to 8,368 civilian casualties from 24 February to 15 May, commenting that the realistic numbers were likely far higher. Its conclusion urged both parties to “respect and ensure respect, at all times and in all circumstances, for international human rights law” \cite{ohchr2022armed}. 
\begin{table}
  \caption{Examples of Telegram Posts. \textit{During the Russia-Ukraine war, Telegram emerged as a significant news source, providing a substantial amount of textual data related to the conflict. We utilized NLP tools to detect mentions of human rights violations within this data.}}
  \label{table:telega}
  \begin{tabular}{p{1.5in}p{1.5in}}
    \toprule
    Telegram Post &  Translation \\
    \midrule
    \begin{otherlanguage*}{russian} `Наш танкист показал, что происходит на Угледарском направлении. Военный вместе со своим экипажем танка укрывается в воронке от снаряда, пока над ними летают украинские мины. 
        `Здесь пц. Но мы давим', – сказал он.'
\end{otherlanguage*} & `Our tanker showed what is happening in the Ugledar direction. A soldier, along with his tank crew, takes cover in a shell crater while Ukrainian mines fly over them. `Here it's a **. But we are pushing,' he said.'   \\
        \\
    \begin{otherlanguage*}{russian}  `Ряд источников сообщает, что в районе Изюма в воздушном бою сбит украинский Су-27.'
\end{otherlanguage*} &  `A number of sources report that a Ukrainian Su-27 was shot down in an air battle near Izyum.'    \\
        \\
    \begin{otherlanguage*}{russian}  `Під завалами школи у Краматорську Донецької області, яку 21 липня зруйнували російські війська, рятувальники виявили тіла трьох загиблих' \end{otherlanguage*} &  `Under the rubble of a school in Kramatorsk, Donetsk region, which was destroyed by Russian troops on July 21, rescuers found the bodies of three victims'\\
        \\
    \begin{otherlanguage*}{russian}  `В результате 14 мирных граждан погибли, 38 человек, в том числе двое несовершеннолетних, получили ранения' \end{otherlanguage*} & `As a result, 14 civilians were killed, 38 people, including two minors, were injured'\\
        \\
  \bottomrule
\end{tabular}
\end{table}

Human rights and its protection are paramount, on both the global and national scale, due to the baseline it establishes among humanity. International efforts dating back centuries, from the Geneva Conventions to the Universal Declaration of Human Rights, stem from an agreement that said rights are fundamental to a “decent life.” A state which values foremost the quality of life of its people, one that is willing to compromise their “absolute sovereignty” for a limited,  but more “responsible” one \cite{marks2016human}, lays a stronger foundation towards a successful society. 

Yet, the patterns seem to indicate a concerning deterioration rather than strengthening of human rights. The World Justice Project, a 16-year-ongoing research effort that calculates countries’ Rule of Law scores based on factors drawing from human rights norms (equal treatment and absence of discrimination, constraints on government powers), reports a concerning recent erosion to the rule of law. Since 2015, the Rule of Law has “weakened in 64\% of countries studied in the WJP Index” and “overall scores have declined by an average of 2.6\% globally.” Authoritarian tendencies, including “diminished accountability and eroded protection of fundamental rights,” are directly cited as a leading cause \cite{wjp2022insight}. 

This trend has not gone unnoticed in the Computer Science field. Efforts towards HRV prevention, both broad and niche, continue to incorporate new technology, from general ethical discussions regarding AI as a potential tool for documenting war crimes \cite{dawes2022weapons} to the much more practical Syria Project and its geographical mapping of decades worth of report-based data. The internet also serves as a paramount assistant in its own right; the OHCHR website makes clear that using their online tool is the “recommended” option for HRV reporting \cite{ohchr2023website}, while the International Criminal Court (ICC) publicly lists an email address to report potential violations \cite{icc2023website}. 

Since the interpretation of human rights can drastically vary depending on culture \cite{bachelet2022atrocity}, the practice of judging international HRVs requires an objective database of evidence, made through the identification, verification, and documentation of HRVs. Specifically regarding the Russia-Ukraine war, the aforementioned OHCHR article places emphasis on “the comprehensive preservation of all types of evidence related to alleged international humanitarian and human rights law violations, including digital evidence.” Interestingly, although not explicit, the article references social media by mentioning “videos apparently recorded by combatants [being] available online” \cite{bachelet2022atrocity}—this is hardly surprising, considering the active precedent for social media content serving as evidence in ICC \cite{irving2017icc} as well as the overall recognition of the evidentiary value said content can have.

In this research, we aim to detect mentions of human rights violations in textual data in the Russian language by utilizing Natural Language Processing (NLP) techniques. The motivation is to help NGOs, governments, and other researchers further the protection of human rights. 

We have chosen to collect our data through the social media platform Telegram due to its close association with the Russia-Ukraine war\footnote{https://www.nytimes.com/2022/04/16/world/europe/russian-propaganda-telegram-ukraine.html}. Foremost, there is the matter of popularity—since the beginning of the war, the application has seen a rapid growth in user base, with the  number of users jumping from 27.5 million in February, 2022  to 40.6 million in March, 2022 \cite{Yuzbekova2022telegram}. Secondly, having been utilized by multiple sources, from government officials on both sides to news channels, the application has rapidly become the center for dispersing politically sensitive information among post-Soviet users. Consequently, we create and label datasets from publicly available Telegram channels that cover political news or purely focus on the Russia-Ukraine war. Examples of posts from Telegram are provided in Table \ref{table:telega}. 

It is a common practice to conduct NLP analysis on Twitter data. However, within the relevant geographical area of post-Soviet countries, Telegram has emerged as a more popular communication medium, with a penetration rate of 64.4\% in 2022, in comparison to Twitter's rate of only 5.7\%.\footnote{per Statista Research Department https://www.statista.com/statistics/867549/top-active-social-media-platforms-in-russia/ Accessed: March 9, 2023} It should also be noted that the Russian government has blocked Twitter, making it inaccessible without a VPN.\footnote{https://www.theguardian.com/world/2022/mar/04/russia-completely-blocks-access-to-facebook-and-twitter} In contrast, Telegram has become a crucial communication and news source for Russian-speaking populations in both Russia and Ukraine, due to easier anonymous and secure information exchange. In Ukraine, the usage of Telegram has increased significantly after the invasion, with a penetration rate of 66\%.\footnote{https://mezha.media/en/2022/08/06/the-most-popular-social-networks-in-ukraine-during-the-war-research-by-global-logic/}

For our task of detecting human rights violations, it is essential to recognize that Telegram is among the most secure social media and messaging platforms prevalent in the region, and thus, it is a safer avenue to report any potential atrocities. Hence, we see Telegram having a strong scientific interest for the purpose of our research.

We hypothesize that an effective machine learning model can be trained on the Telegram textual data for the task of detecting Human Rights Violations.

Additionally, in this paper we introduce two separate datasets:
\begin{enumerate}
    \item A large corpus with 2.3 million Telegram posts in Russian and Ukrainian.
    \item  A subset of (1), annotated to indicate HRV, that contains 3,076 posts that were split into 15,693 sentences. Annotation is binary (HRV versus non-HRV) and on a sentence level.
\end{enumerate}

\textbf{We summarize the key novelty of our contributions as follows:}
\begin{itemize}
    \item This is the first research that utilizes Machine Learning (ML) to automatically detect human rights violation in Russian language; 
    \item We release a corpus containing 2.3 million Russian and \\Ukrainian Telegram posts, many regarding the war;
    \item We publicly release a dataset labelled to indicate mentions of human rights violations in Russian and Ukrainian languages.
\end{itemize}

We believe that in the future this research can be incorporated into tools that advance UN Sustainable Development Goals by promoting peaceful societies for sustainable development.

\section{Literature Review}

\subsubsection{Human Rights Violation and Machine Learning}

With social media becoming a strongly entwined part of everybody’s life, we receive a massive amount of data. While this data can provide insights on many issues, we explore the possibility and previous attempts to use the social media data to detect possible instances of human rights violations.  Multiple researchers attempted to use computational methods to help with human rights violations detection. 

Some research in this field involves using internet image processing where instances of violations can be observed \cite{kalliatakis2017detection} reaching mean average precision of 88.1\%.  The authors' focus was on four HRVs: `Child Labour', `Police Violence', `Refugees', and `Child Soldiers'.  While computer vision is a promising area, we aim to focus only on the textual data. Hence, we look closer at the available approaches that involve processing textual data.  Among those, for example, there is a study where the authors use statistical methods such as non-parametric heterogeneous social media graphs to identify HRVs on Twitter \cite{chen2015human}. The researchers were able to identify clusters of tweets that were relevant to certain HRV related events in Mexico. This work involved extracting tweets based on particular keywords. Another approach explores NLP techniques to detect indications of human rights violations in Arabic Twitter \cite{alhelbawy2020nlp}. The scientists applied deep neural networks (CNN, LSTM, and bidirectional LSTM) to perform a binary classification (human rights abuse (HRA) or non-HRA).  The researchers define HRA as `crimes committed by government, militia, or organisations against civilians' (\citeyear{alhelbawy2020nlp}).  This approach yielded a precision close to  85\% and an $F_1$ score of 75\%. (While similar to our task, in this study the dataset was extracted from Arabic Twitter using 237 violence indicating keywords hence changing the data natural distribution. In this study we make our task more challenging by randomly extracting posts). We found this approach appealing yet would expect to see transformers being used as a possible model as well. While we see a neural network approach for text classification as state of the art, there is work on improving those results with deep learning ensemble methods, such as in Zimmerman et al. that were able to increase the $F_1$ score by an average of 2\% in a hate speech detection task (\citeyear{zimmerman2018improving}).

\subsubsection{Related Tasks}

Since human rights violation detection using NLP techniques is a relatively new niche, we also look at closely related research, such as hate speech and abusive language detection. For example, Gitari et al.(\citeyear{gitari2015lexicon}) employed sentiment analysis in a form of subjectivity detection to create a hate speech classifier. In their survey review, Schmidt and Wiegand (\citeyear{schmidt2017survey}) present a vast variety of approaches existing to detect hate speech using natural language processing. They point to the following types of features that can be useful for a classification: simple surface features (bag-of-words), word generalization, sentiment analysis, lexical resources, linguistic features, knowledge-based features, meta information, and multimodal information (images, video).  Although human rights violation detection differs from detecting hate speech, deriving classification methods from the field of hate speech recognition could be valuable. 

\subsubsection{Telegram Data in Research}

Another pillar of this paper is using data from Telegram, a widely used social media  in post-Soviet regions, which is more appropriate for our purposes.  Yet, research with Telegram data is sparse and none of it is based on Russian language posts. Some of the few existing papers that use Telegram data explore extreme actors. For instance, Urman and Katz [\citeyear{urman2022they}] explore far-right political networks and their pattern on Telegram.

Similarly, Rogers [\citeyear{rogers2020deplatforming}] explored Telegram data in order to detect and remove extreme actors from the platform, Walther and McCoy [\citeyear{walther2021us}] study U.S. extremism in Telegram, Akbari and Gabdulhakov [\citeyear{akbari2019platform}] analyzed Telegram as a tool of resistence in Russia and Iran. Another group of researchers are focused on the analysis of the quality of education via Telegram \cite{abu2020telegram,singh2020rethinking,alakrash2020effectiveness,citrawati2021telegram}.

\subsubsection{Russia-Ukraine War and NLP}
A few efforts have been made to analyze Twitter data concerning the Russia-Ukraine conflict, all utilizing English language tweets. For instance, Thapa et al. created a multimodal dataset from English tweets that were annotated as either hate speech or not, specifically focusing on the Russia-Ukraine war \cite{thapa2022multi}. Various research groups have also attempted sentiment analysis on English Twitter data related to the Russia-Ukraine conflict \cite{angdreseydecision,shlkamyrussia}. A recent study explored the perceived public opinion regarding the war by building on Twitter data, also strictly in English \cite{mir2023exploring}.

In another study, Donofrio et al. performed an exploratory analysis of the Twitter audience of the Ukrainian and Russian government profiles [\citeyear{donofrio2023russia}].

To the best of our knowledge, our study is the first attempt to use Telegram data regarding the Russia-Ukraine war to detect possible human rights violations in Russian language. We believe that this is an important task that can possibly help NGOs and governments to keep track of possible HRV and war crimes.

\section{Data}
For the present study, we aimed to identify instances of human rights violations (HRVs) in textual data from public Telegram channels covering the Russia-Ukraine war and political news. To achieve this goal, we first identified a total of 95 public channels covering the aforementioned topics. These channels encompassed a wide range of pro-Russian and pro-Ukrainian affiliations, ensuring a diverse range of perspectives. Examples of Telegram posts with their English translations can be found in Table\ref{table:telega}.

\begin{table}
    \centering
    \begin{tabular}{lr}
        \hline
        \hline
        Total Number of Channels & 95     \\
        Channels Leaning to Russia & 28    \\
        Channels Leaning to Ukraine & 26    \\
        Channels with Unclear Affiliation & 41    \\
        Average Number of Posts per Channel & 24,300     \\
        \hline
    \end{tabular}
    \caption{Corpus Description}
    \label{table:corpus}
\end{table}

We collected all posts from each channel starting from the beginning of the war on February 24th, 2022 until September 11th, 2022. This yielded a large corpus consisting of posts written in both Russian and Ukrainian languages.  We provide descriptive data for the corpus in Table\ref{table:corpus}. We openly release this unique corpus for scientific community use.\footnote{https://github.com/author-of-the-paper} 

\subsection{Annotated Data}

To obtain a representative sample of posts related to the Russia-Ukraine conflict, we randomly extracted 30 posts from each of the 95 public channels identified on Telegram. This approach was chosen to ensure the sample's representativeness.

We split the posts into sentences for our task utilizing NLTK's PunktSentenceTokenizer library for Russian.\footnote{https://github.com/Mottl/ru\_punkt} The decision to partition the posts into sentences was made with the intention of conducting a more nuanced analysis. Additionally, since certain posts may contain numerous HRVs, our aim was to effectively capture each instance. The average length of a sentence in our annotated dataset was 13 words, with a mode of 7 words.  

For the labeling task we employed the services of four annotators who were native Russian speakers and possessed diverse political affiliations. The annotators were provided with instructions in the Russian language. 

\begin{table}
    \centering
    \begin{tabular}{p{1.5in}p{1.5in}}
        \hline
        Positive examples  & Translation \\
        \hline
        \begin{otherlanguage*}{russian} По предварительным данным, жертвами обстрела стали 9 мирных жителей (8 погибли и 1 ранен)», — сообщает СЦКК ЛНР\end{otherlanguage*} & According to preliminary data, 9 civilians became victims of the shelling (8 were killed and 1 wounded),” the LPR JCCC reports. \\
        \hline
        \begin{otherlanguage*}{russian} По его словам, в результате ночного ракетного удара по Запорожью пострадали многоквартирные дома и улицы частного сектора города.\end{otherlanguage*} & According to him, as a result of the night missile attack on Zaporozhye, apartment buildings and streets of the private sector of the city were damaged. \\
        \hline
        Negative examples  & Translation \\
        \hline
        \begin{otherlanguage*}{russian} ВСУ официально заявили об уничтожении 44 из 50 запущенных по Украине ракет.\end{otherlanguage*} & The Armed Forces of Ukraine officially announced the destruction of 44 out of 50 missiles launched in Ukraine. \\
        \hline
        \begin{otherlanguage*}{russian} О какой победе...Неужели украинцы не понимают, что их используют и всем наплевать на их жизни?\end{otherlanguage*} & About what a victory ... Do the Ukrainians really not understand that they are being used and that no one cares about their lives? \\
        \hline
    \end{tabular}
    \caption{Examples of Labeled Sentences: (HRV or `1' and non-HRV or `0'}
    \label{tab:lab}
\end{table}

Each post in our dataset was associated with the following information: date and timestamp, post ID, channel ID, and message.  Human annotators labeled the dataset using Label Studio\footnote{https://labelstud.io/} software and Google spreadsheets.

The human annotators tasked with labeling the data were instructed to identify sentences that potentially mention one or more  human rights violations (HRVs). Specifically, we restricted this project to the following types of HRVs: killing civilians, destruction of civil objects, rape, torture, execution, or mistreatment of prisoners. Out of the The Universal Declaration of Human Rights\footnote{https://www.ohchr.org/en/what-are-human-rights/international-bill-human-rights}, we narrowed down the list of HRV to just six in order to make the task more specific for the annotators. We chose these particular  six as the most relevant to the war and, hence, more likely appearing in our dataset. (We grouped `rape, torture, execution' as one label, and hence have four types of HRVs in total). The final labels were binary in nature: 1 indicates the presence of a possible HRV, while 0 indicates the absence of HRVs. The annotators were presented with the full posts, but labeled the data on a sentence level, thereby facilitating more detailed and nuanced future analysis. We provide the annotator guidelines (in Russian) in our Github repository.\footnote{The detailed annotation guidelines are available at [[url withheld for anonymization]].}

Most of the core annotations were performed by two annotators and adjudicated by a third. Specifically, a first annotator pre-labeled all 15,693 sentences (3076 posts); 228 sentences were labeled as HRV (positive class) and 15,465 as non-HRV (negative class). Then three subsets of posts were extracted: (1) those containing at least one sentence pre-labeled as positive, (2) an equal-sized random sample of posts where all of the sentences had been pre-labeled negative, and (3) all additional posts containing at least one sentence labeled positive by a preliminary version of our classifier (details of our model are covered in Section \ref{Methods}). These posts were presented to the two annotators in a random order and double annotated. All disagreements were adjudicated by a different analyst who is a Russian speaking political expert. All the other pre-labeled posts were left in the dataset with their negative (non-HRV) sentence labels. After the adjudication number of positive examples changed to 273 and negatives  - to 15,410.

The Cohen's Kappa score \cite{cohen1968weighted}, which is commonly used to evaluate annotator agreement, was computed to assess the level of agreement between the two independent annotators. The Cohen's Kappa score obtained indicates moderate agreement according to Landis and Koch [\citeyear{landis1977measurement}]. This score suggests that the task of identifying instances of human rights violations in social media posts is challenging, and different annotators may have varying perceptions regarding what constitutes such violations. 
We provide positive and negative labeled examples in Table \ref{tab:lab}.

The final size of this annotated dataset is 15,693 labeled sentences: 273 examples of HRV class and 15,420 of non-HRV labeles examples. 

\subsection{Data Partitioning}
From the annotated dataset previously described we randomly partitioned 80\% of the posts for use as a training set, while reserving 20\% for a held-out test set, detailed in \ref{test_data}.
A description of our annotated training and test data is provided in Table \ref{table:final_data}.

\subsection{Training Data}\label{keywords}

\begin{figure}
\centering
\includegraphics[scale=0.4]{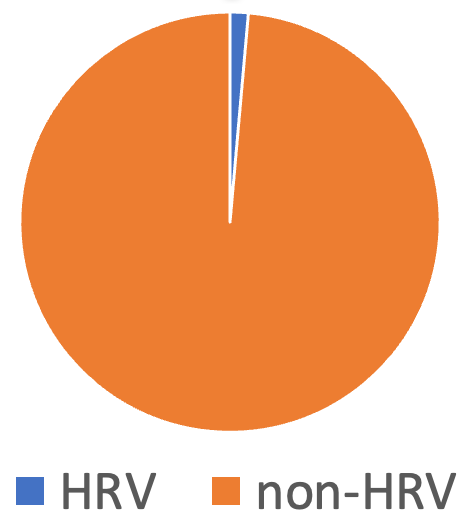} 
\caption{Class Imbalance}
\label{fig:wrapfig}
\end{figure}

As described earlier and illustrated in Figure \ref{fig:wrapfig}, our training data is highly imbalanced with just around 2\% of sentences being labeled as HRV. We attempted to mitigate this class imbalance by utilizing the following data augmentation techniques: Back Translation and ChatGPT Data generation.

\textbf{Back Translation}. We performed paraphrasing of the positive examples through  translation chains such as Russian-Arabic-Japanese-Russian \cite{beddiar2021data}. 
This approach yielded another 228 HRV training examples per chain. In total, we utilized 5 different translation chains, each involving two languages in addition to the Russian source language (see Table \ref{table:bt}). This resulted in a total of 1,140 additional positive examples to augment the training dataset, bringing number of positive examples to be close to 10\% of the training data.

\begin{table}
    \centering
    \begin{tabular}{lr}
        \hline
      
        Iteration & Language Chain     \\
        \hline
        \\
        1 & Russian-Arabic-Japanese-Russian    \\
        2 &  Russian-Turkish-Farsi-Russian    \\
        3 & Russian-Chinese-Polish-Russian    \\
        4 & Russian-Hebrew-German-Russian     \\
        5 & Russian-Thai-Greek-Russian     \\
        \hline
    \end{tabular}
    \caption{Back Translation Language Chains}
    \label{table:bt}
\end{table}

\textbf{Chat GPT Data Generation}. Furthermore, in order to enhance the training dataset, we implemented a method of generating synthetic data through the utilization of ChatGPT\footnote{https://openai.com/blog/chatgpt}, which is a sizeable language model capable of generating language autonomously. Our approach entailed prompting ChatGPT to generate sentences in the Russian language that depict violations of human rights. Through a process of prompt engineering, we selected two prompts that yielded the most appropriate data for our purposes.

The first prompt instructed ChatGPT to produce sentences in the Russian language that exemplify instances of human rights violations during the Russia-Ukraine conflict. It is noteworthy that the initial output generated by ChatGPT exhibited a bias towards violations committed exclusively by the Russian side. This bias was subsequently addressed by posing the question to Chat GPT, "Why are all the violations presented as being committed by Russia only?" This led to a more balanced output of data.

Prompt 2 instructed ChatGPT to act as a Telegram channel and generate ten posts, each comprising a single sentence, that describe anecdotes of human rights violations in the context of the Russia-Ukraine war in 2022. This scenario-based prompt produced examples that were more aligned with the intended objective. However, during the data retrieval phase utilizing ChatGPT, we observed the model's instability, whereby the same prompt could elicit either a refusal to respond due to ethical considerations or a range of responses with varying styles.

After we reviewed the examples, five of them was removed as being irrelevant. We generated a total of 510 positive examples with this approach and included them as a part of our training data.

\textbf{Sets of Training Data.}
\label{train_data}
We ran our models on the following sets of training data:
\begin{itemize}
    \item The Base Training Dataset (D1): 80\% of the annotated data (2,466 posts, 12,554 sentences).
    \item Augmented Training Dataset (D2): The Base Training Dataset + Augmented Data using Back Translation (3,606 posts, 13,694 sentences).
    \item Dataset 3 (D3): The Base Training Dataset + ChatGPT data (13,064 sentences).
    \item Dataset 4 (D4): Augmented Training Dataset + ChatGPT data (14,429 sentences).
\end{itemize}

\subsection{Testing Data}
\label{test_data}

The testing dataset consisted of the remaining 20\% of the annotated posts, resulting in a size of 610 posts or 3,139 sentences.

\begin{table}
    \centering
    \begin{tabular}{lrr}
        \hline
         Parameter & Train Data & Test Data    \\
        \hline
        \\
        Total Number of Posts & 2,466  &  610 \\
        Total Number of Sentences & 12,554  & 3,139\\
        Number of HRV Sentences & 228 & 45   \\
        Number of non-HRV Sentences & 12,326 &  3,094 \\
        Number of HRV Posts & 90 &  25  \\
        Number of non-HRV Posts & 2,376 &  585 \\
        \hline
    \end{tabular}
    \caption{The Base Dataset Description}
    \label{table:final_data}
\end{table}

\section{Methods}

The aim of our study is to detect textual mentions of human rights violations (HRV) in Telegram posts. In an application, these full posts would be forwarded, with the detected mentions highlighted, to an analyst for review, archival and action. Therefore, our data is labeled at the sentence level, rather than post level, to allow the sentence-level mentions to be emphasized. Hence, the methods here focus on performing classification at the sentence level.\footnote{Similar $F_1$ scores were seen when either classifying at the post level or rolling the sentence-level classifications up to evaluate at the post level. However, post-level classification results have higher variance, since the number of positive test examples is roughly cut in half.} 

In addition, our emphasis is on identifying most potential references to HRV, at the cost of a higher false positive rate. Therefore, we prioritize evaluating our system's recall over precision. Accordingly, we have opted to employ the $F beta$ score, where $beta$ is set to 2 ($F_2$), as our primary evaluation metric \cite{rijsbergen1979information}. This measure enables us to place greater emphasis on recall than precision in our evaluation. We also report $F_1$ scores.

\subsection{Baseline} 
In order to evaluate how well this task can be performed using a straight forward method, we evaluated a keyword baseline approach. 


We utilized the \textit{yake} library\footnote{https://pypi.org/project/yake/} to extract 30 keywords and presented both the keywords themselves and their corresponding English translations in a table \ref{table:keyw}.
  
  \begin{table}
  \tiny
  \label{table:keyw}
  \centering
  \begin{tabular}{ p{1.5cm}p{1.5cm}||p{1.5cm}p{1.5cm}}
    \toprule
    Keyword &  Translation & Keyword &  Translation\\
    \midrule
    \begin{otherlanguage*}{russian}`результате' \end{otherlanguage*} & `resulted' &\begin{otherlanguage*}{russian}`людей'  \end{otherlanguage*} & `people'\\
    \begin{otherlanguage*}{russian}`граждан' \end{otherlanguage*} & `civilians' &\begin{otherlanguage*}{russian}`Буча'  \end{otherlanguage*} & `Bucha'\\
     \begin{otherlanguage*}{russian}`пытки' \end{otherlanguage*} & `tortures' &\begin{otherlanguage*}{russian}`мирного'  \end{otherlanguage*} & `peaceful'\\
     \begin{otherlanguage*}{russian}`города' \end{otherlanguage*} & `city' &\begin{otherlanguage*}{russian}`погиб'  \end{otherlanguage*} & `died'\\
     \begin{otherlanguage*}{russian}`дома' \end{otherlanguage*} & `houses' &\begin{otherlanguage*}{russian} `жителей'  \end{otherlanguage*} & `inhabitants'\\
     \begin{otherlanguage*}{russian}`мирных' \end{otherlanguage*} & `of peaceful' &\begin{otherlanguage*}{russian} `разрушен'  \end{otherlanguage*} & `destroyed'\\
    \begin{otherlanguage*}{russian}`российские' \end{otherlanguage*} & `Russian' &\begin{otherlanguage*}{russian} `обстрела'  \end{otherlanguage*} & `shelling'\\
    \begin{otherlanguage*}{russian}`ВСУ' \end{otherlanguage*} & `AFU'\footnote{Armed Forces of Ukraine} &\begin{otherlanguage*}{russian} `Информации'  \end{otherlanguage*} & `Information'\\
    \begin{otherlanguage*}{russian}`Окупанти' \end{otherlanguage*} &   `Invaders' &\begin{otherlanguage*}{russian} `получили'  \end{otherlanguage*} & `received'\\
    \begin{otherlanguage*}{russian} `ранения' \end{otherlanguage*} & `injuries' &\begin{otherlanguage*}{russian} `детей'  \end{otherlanguage*} & `children'\\
    \begin{otherlanguage*}{russian} `пострадали' \end{otherlanguage*} & `injured' &\begin{otherlanguage*}{russian} `Мариуполя'  \end{otherlanguage*} & `Mariupol'\\
    \begin{otherlanguage*}{russian} `Милостив' \end{otherlanguage*} & `Milostiv' &\begin{otherlanguage*}{russian} `САУ' \end{otherlanguage*} & `SAU'\\
    \begin{otherlanguage*}{russian} `удар' \end{otherlanguage*} & `strike' &\begin{otherlanguage*}{russian} `сообщает' \end{otherlanguage*} & `informs'\\
    \begin{otherlanguage*}{russian} `дітей' \end{otherlanguage*} & `children' &\begin{otherlanguage*}{russian} `словам' \end{otherlanguage*} & `words'\\
     \begin{otherlanguage*}{russian} `видео' \end{otherlanguage*} & `video' &\begin{otherlanguage*}{russian} `человека' \end{otherlanguage*} & `person'\\

    \bottomrule
   
\end{tabular}
\caption{Keywords}
\end{table}

  It is worth noting that we also considered the suitability of the nltk\footnote{https://www.nltk.org/} and spaCy\footnote{https://spacy.io/}  Python libraries for the purposes of automated keyword extraction. However, based on our experimentation with a Russian dataset, we observed that the \textit{yake} library produced a more pertinent set of keywords. Consequently, we adopted the \textit{yake} library as our chosen method for keyword extraction.

Subsequently, we designated all sentences that contained at least one of the keywords extracted from the list as representing human rights violation (HRV), whereas those lacking any such keywords were categorized as non-HRV.

\subsection{Models}
 For our tasks, we chose to test transformers as they are the state-of-the-art models for textual data.\footnote{https://gluebenchmark.com/leaderboard} First, we employed  a base multilingual BERT model (\textbf{mBERT})  which is a pre-trained deep learning language model with 12 layers (Tranformer blocks), a word embedding (hidden) size of 768, and 12 self-attention heads (110M total parameters) \cite{devlin2018bert}.  Specifically, we used the Huggingface bert-base-multilingual-uncased model.\footnote{https://huggingface.co/bert-base-multilingual-uncased}

We also evaluated \textbf{XLM-RoBERTa}\footnote{https://huggingface.co/xlm-roberta-base}, a transformer model pre-trained on a large corpus in a self-supervised fashion on 2.5TB of filtered CommonCrawl data spanning 100 languages  \cite{conneau2019unsupervised}. 

Finally, we also tried \textbf{RuBERT}, a BERT-based model specifically trained on the Russian language.\footnote{https://huggingface.co/DeepPavlov/rubert-base-cased} RuBERT is a monolingual transformer-based language model that was trained from multilingual BERT with the addition of extra Russian vocabulary tokens \cite{kuratov2019adaptation}. 

We conducted $k$-fold cross-validation \cite{kohavi1995study} with k set to 5 for hyper-parameter tuning and model validation. During the tuning phase, we selected the training dataset, model, number of training epochs, and the block fine-tuning strategy that exhibited the optimal performance. Following cross-validation on the training dataset, we determined that a total of 5 epochs was optimal. Additionally, we elected to freeze the first six model blocks and fine-tune only the final six blocks as a means of mitigating overfitting and enhancing model generalization. 


Several parameters were not tuned in the experiment. Specifically, all models were trained utilizing a categorical cross-entropy loss function, the AdamW optimizer  \cite{loshchilov2017decoupled} and a learning rate of 5e-5. In light of the class imbalance observed in the dataset, the weights in the loss function were modified to 0.17 for the non-HRV class and 0.83 for the HRV class. Alternative weight configurations of [0.5, 0.5] and [0.2, 0.8] were explored. 
The most optimal results for the experiment were determined empirically to be associated with the reported combined weight configuration.

\section{Results}
As previously stated, our emphasis lies on recall, therefore, we have adopted the utilization of $F_2$ as a principal evaluation metric. The most proficient model in our study, namely mBERT trained on the Augmented Dataset, obtained an $F_2$ score of 0.71, with precision and recall rates of 0.47 and 0.82, respectively. We observed that the incorporation of back translation data resulted in an enhancement of 0.38 points in the performance of mBERT relative to training only on the Base Training Dataset. All the performance scores are provided in Table \ref{tab:f1}.

\begin{table}[t]
    \centering
    \begin{tabular}{lllll}
        \hline
        
        \hline
        Approach and Training Data & $P$ & $R$ & $F_1$ & $F_2$ \\
        \hline
        \\
        \textit{D1:Base Training Dataset } & &  \\
        \\
        Keywords Based (baseline)   &0.12 &  0.65 & 0.19 & 0.35 \\
        \\
        mBERT base     & 0.42 & 0.31 & 0.36 & 0.33 \\
        RuBERT   & 0.46  &  0.36  & 0.40 & 0.38 \\
        XLM-RoBERTa & 0.38  &  0.33  & 0.36 & 0.34  \\
        \hline
        
        \textit{D2: Augmented Dataset} \\

        \\
        mBERT base     & 0.47 & 0.82 & 0.60 & \textbf{0.71}\\
        RuBERT   & 0.45  &  0.80  & 0.58 & 0.69\\
        XLM-RoBERTa & 0.41  & 0.80  & 0.54 & 0.67 \\
        
        \hline
  
        \textit{D3: Base Training Dataset} \\
        \textit{+ ChatGPT data}    \\
        \\

        mBERT base     & 0.45 & 0.40 & 0.42 & 0.41\\
        RuBERT   & 0.54  &  0.31  & 0.39 & 0.34\\
        XLM-RoBERTa & 0.39  & 0.53  & 0.45 & 0.49 \\
        \hline

        \textit{D4: Base Training Dataset} \\
        \textit{+ ChatGPT data }\\
        \textit{+ Augmented Data}  \\
        \\

        mBERT base     & 0.39 & 0.82 & 0.53 & 0.67 \\
        RuBERT   & 0.43  &  0.69  & 0.53 & 0.62\\
        XLM-RoBERTa & 0.35  & \textbf{0.87}  & 0.50 & 0.67\\
        \hline
        \hline
    \end{tabular}
    \caption{Test Set Performance: Breakdown per model and training dataset.}
    \label{tab:f1}
\end{table}

\section{Discussion}

\textbf{Model Performance.} 
All of the models we tried performed best with the Augmented Dataset (D2). We attribute that to the fact that there is a substantial class imbalance in the Base Training Dataset. The top performing model is multilingual BERT, while RuBERT consistently performed just slightly worse (dropping 0.02 to 0.07 in the $F_2$ score) with the exception of training on D1 where RuBERT scored 0.05 points higher. This is counter-intuitive because both models have the same BERT base architecture and RuBERT was specifically trained for Russian. The difference appears to be in the training data. One possible explanation is that, as a result of the multi-language translation chain incorporating languages from divergent groups, the final back-translated ``Russian" language resembles these intermediary languages. This could result in a lower performance for the Russian-specific LM, RuBERT, while mBert could be robust to this grammatical and stylistic variation given its multi-lingual pre-training corpus.

XLM-RoBERTa achieved the highest $F_2$ score when trained on the augmented dataset (D2). Notably, compared to other models, XLM-RoBERTa demonstrated the most significant improvement in performance when trained on the D3 dataset. This observation implies that the inclusion of ChatGPT had the greatest impact on this particular model, resulting in a 15\% increase in $F_2$ score.

$F_2$ score for all models evaluated. Notably, incorporating ChatGPT data resulted in an 8\% increase in mBERT and a 15\% increase in XLM-RoBERTa, compared to training solely on the Annotated Dataset. However, for RuBERT, the addition of ChatGPT resulted in a 4\% decrease in performance.

The study found that combining all training datasets (D4) did not result in any performance improvement for the models evaluated, compared to the results achieved with D2 alone. Furthermore, the precision of the models was significantly compromised.

\textbf{Class Imbalance.} It is important to acknowledge the challenge that arises from the imbalance in the training data set, which includes only a scant number of instances of HRV compared to the considerable number of non-HRV sentences. Despite this discrepancy, our model was able to attain an $F_2$ score of 0.71. Nonetheless, the recall value of 0.84 and the precision of 0.45 suggest that the model is able to relatively well identify relevant items whenever they are present. Nevertheless, only 45\% of the extracted items are deemed relevant. We will address the trade-off between precision and recall for our study in subsequent paragraphs.

Our attempts to tackle this challenge through data augmentation and assigning specific weights in the loss function have led to marginal improvements. Hence, the class imbalance remains a persistent obstacle in our data. Also, we recognize that the biggest issue to tackle now is an sub-optimal precision.

\textbf{Impact of Data Augmentation.} The issue of class imbalances was partially mitigated by the acquisition of additional positive examples through back translation. As demonstrated in Table  \ref{tab:f1}, performance of all the models was observed to improve upon the inclusion of augmented back translation data. Specifically, the mBERT trained solely on the Base Training Dataset yielded an $F_2$ score of 0.33, while augmenting the training data resulted in an increase in the $F_2$ score to 0.71. It is noteworthy that the use of the Augmented Dataset for training resulted in a recall improvement of more than twofold across all models.

The utilization of a ChatGPT approach for augmenting the training data yielded enhanced performance compared to training solely on the Base Training Dataset (D1). XLM-RoBERTa exhibited the most substantial improvement with a 15\% increase in the $F_2$ metric, alongside a marginal enhancement in precision by 1\%, and a considerable improvement in recall by 20\%.

Despite the fact that in our study, back translation brought the primary benefit,   we note that data augmentation via ChatGPT also improved the performance. Our study focused predominantly on the Russian language, which is a high-resource language that can be translated accurately and with ease. Nonetheless, for other languages that do not possess such resources, data augmentation via ChatGPT represents a viable alternative for improving results.

\textbf{Recall vs. Precision.} Our system's primary objective is to identify the maximum number of instances of HRV, even at the expense of erroneously classifying some non-HRV items. As such, our focus is on maximizing recall, rather than precision. (With that understanding, we utilize the $F_2$ metric as our principal evaluation criterion to evaluate our system's performance.)

Our analysis of the various combinations of models and training data has revealed notable performance disparities between the most effective options, such as mBERT and RuBERT on both the Augmented Training Dataset and the Base Training Dataset. Specifically, we observed that the most effective model was mBERT trained on the Augmented Training Dataset, which yielded a high recall rate of 0.82, albeit at the cost of a lower precision of 0.47 (and an $F_2$ score of 0.71). In contrast, RuBERT trained on the Augmented Training Dataset exhibited slightly inferior performance, with an $F_2$ score of 0.69, despite an identical recall rate of 0.80, but with a lower precision of 0.45.

Remarkably, we observed the highest recall rate of 0.87 when training XLM-RoBERTa on the combined dataset (D4), while simultaneously maintaining an $F_2$ score of 0.67.

\textbf{The Cohen's Kappa score.} Despite the moderate value of the Cohen's Kappa score, it is crucial to underscore the significance and complexity of our undertaking. We remain optimistic that our datasets and preliminary systems will provide valuable groundwork for further research by other scholars.

\textbf{Error Analysis.} 
It appears that the easiest type of HRV for the system to detect is `killing of civilians', as seen in the following example, \begin{otherlanguage*}{russian} `Вчера вечером вблизи своего дома в Северодонецке из-за вражеских обстрелов погибли мужчина и женщина, еще две женщины убиты российскими обстрелами в Лисичанске и Привилле.' \end{otherlanguage*} (`Last night, a man and a woman were killed near their home in Severodonetsk due to enemy shelling, two more women were killed by Russian shelling in Lisichansk and Priville').  

The most common \textit{false negative} error is associated with `Civil Objects Destruction'. Examples of this error are \begin{otherlanguage*}{russian} `Загарбники здійснили масований обстріл міста Слов’янськ Донецької області — у зону обстрілу потрапив місцевий ринок.' \end{otherlanguage*} (`The invaders launched a massive shelling of the city of Slovyansk, Donetsk region — a local market fell into the shelling zone.') and \begin{otherlanguage*}{russian} `Когда увидел своими глазами, ужаснулся: два десятка домов в хлам' \end{otherlanguage*} (`When I saw with my own eyes, I was horrified: two dozen houses got destroyed  to the trash'). This errors suggest that adding context into the model may help the model to learn better and avoid similar errors. 

The prevalent \textit{false positive} examples are related to death or other violence but not in a context of HRV. For example: \begin{otherlanguage*}{russian} `Тут можно найти и конституционную реформу 2020 года, и проблемы с независимостью судов, и нарушения права на жизнь в ходе вооруженных конфликтов, и пытки, и домашнее насилие' \end{otherlanguage*} (`Here you can find the constitutional reform of 2020, and problems with the independence of the judiciary, and violations of the right to life during armed conflicts, and torture, and domestic violence'). This suggests that we need to provide more negative examples that include references to death yet are not considered as an HRV.

\textbf{Possible Improvements.} A potential avenue for future improvement could involve integrating contextual information from posts into the classifier. We posit that such an approach has the potential to enhance the performance of the classifier.

\textbf{Future Work.}  With our encouraging results, we are eager to explore further improvements in our system.  
Our research findings offer the possibility of extending this project by developing a multiclass classifier capable of detecting specific types of HRV. We are particularly interested in the four types of HRV mentioned in the annotation description, namely, (1) killing civilians, (2) destruction of civil objects, (3) rape, torture, and execution, and (4) mistreatment of prisoners.

We see our work as a first step in a project to build a more general social media based tool for documenting human rights violations during wars. Specifically, future work includes extrapolating this model to other languages and geographic locations.



\section{Conclusion}

The detection and documentation of human rights violations is a critical task that can have far-reaching implications for individuals, communities, and society as a whole. Despite the importance of this work, identifying and tracking instances of human rights violations can be a challenging and resource-intensive task.

Our study contributes to the limited body of research conducted by machine learning scholars who endeavor to develop an automated, cost-effective means of identifying instances of human rights violations.

In recent years, social media platforms have emerged as an important source of information for detecting and documenting human rights violations. However, the sheer volume of data available on these platforms can make it difficult to separate relevant information from noise. This is especially true in regions of conflict or political unrest, where the spread of misinformation and propaganda is common.

In response to these challenges, researchers have begun exploring the use of machine learning and natural language processing techniques to aid in the detection of human rights abuses on social media. In this study we sought to apply these methods to the Russian and Ukrainian languages, with the aim of identifying potential human rights violations related to the ongoing conflict between the two countries.

To conduct our analysis, we collected and labeled data from 95 public Telegram channels that covered news related to the war. After annotating the dataset and applying data augmentation, we employed classification methods using several transformer models. The best performing model, fine-tuned mBERT, yielded an $F_2$ score of 0.71.

To the best of our knowledge, this study is the first research attempting to use a dataset in Russian language for HRV detection.

Additionally, we believe into importance of join research forces and release for public two new datasets of Telegram posts: (1) large corpus with over 2.3 millions posts and (2) an annotated (HRV/non-HRV) Telegram dataset.

It is our belief that performing analysis on Telegram data represents a distinctive and possibly more informative approach for the Eastern-European region. We aspire to encourage the research community with an interest in this subject matter to leverage our datasets for their investigations.

Overall, we believe this research holds promise for aiding NGOs, governments, and other organizations in the identification and documentation of possible human rights violations on social media. 

\begin{acks}
We appreciate the annotators who volunteered to annotate the data for this project. 
\end{acks}

\bibliographystyle{ACM-Reference-Format}
\bibliography{sample-base}

\appendix

\end{document}